\documentclass[fleqn,usenatbib]{mnras}

\usepackage{newtxtext,newtxmath}
\usepackage[T1]{fontenc}
\usepackage{multicol}
\usepackage{wrapfig}
\DeclareRobustCommand{\VAN}[3]{#2}
\let\VANthebibliography\thebibliography
\def\thebibliography{\DeclareRobustCommand{\VAN}[3]{##3}\VANthebibliography}

\usepackage{graphicx}	% Including figure files
\usepackage{amsmath}	% Advanced maths commands

\usepackage{amssymb}	% Extra maths symbols
\usepackage{subfigure}

\title[Non-axisymmetric structured jet]{Characteristics of gamma-ray burst afterglows in the context of non-axisymmetric structured jets}

\author[J.-D. Li et al.]{
Jin-Da Li$^{1,2}$,
He Gao$^{1,2}$\thanks{E-mail: gaohe@bnu.edu.cn},
Shunke Ai$^{3}$\thanks{E-mail: shunke.ai@whu.edu.cn}
and Wei-Hua Lei$^{4}$
\\
$^{1}$Institute for Frontier in Astronomy and Astrophysics, Beijing Normal University, Beijing 102206, China;\\
$^{2}$Department of Astronomy, Beijing Normal University, Beijing 100875, China;\\
$^{3}$Department of Astronomy, School of Physics and Technology, Wuhan University, Wuhan 430072, China;\\
$^{4}$Department of Astronomy, School of Physics, Huazhong University of Science and Technology, Wuhan, Hubei 430074, China
}

\begin{document}
\label{firstpage}
\pagerange{\pageref{firstpage}--\pageref{lastpage}}
\maketitle

\begin{abstract}

As the most energetic explosions in the universe, gamma-ray bursts (GRBs) are commonly believed to be generated by relativistic jets. Recent observational evidence suggests that the jets producing GRBs are likely to have a structured nature. Some studies have suggested that non-axisymmetric structured jets may be formed through internal non-uniform magnetic dissipation processes or the precession of the central engine. In this study, we analyze the potential characteristics of GRB afterglows within the framework of non-axisymmetric structured jets. We simplify the profile of the asymmetric jet as a step function of the azimuth angle, dividing the entire jet into individual elements. By considering specific cases, we demonstrate that the velocity, energy, and line-of-sight direction of each jet element can greatly affect the behavior of the overall light curve. The radiative contributions from multiple elements may lead to the appearance of multiple distinct peaks or plateaus in the light curve. Furthermore, fluctuations in the rising and declining segments of each peak can be observed. These findings establish a theoretical foundation for future investigations into the structural characteristics of GRBs by leveraging GRB afterglow data.

\end{abstract}


%%%%%%%%%%%%%%%%%%%%%%%%%%%%%%%%%%%%%%%%%%%%%%%%%%

%%%%%%%%%%%%%%%%% BODY OF PAPER %%%%%%%%%%%%%%%%%%

\section{Introduction}

Gamma-ray Bursts (GRBs) are astrophysical phenomena that exhibit an immediate and intense release of gamma-ray radiation from a precise location in the sky, succeeded by a rapid decrease. The prompt emission of GRBs takes place over a span of 0.1-1000 seconds, followed by a multi-wavelength afterglow that can last for months to years \cite[for a review about GRBs]{2018pgrb.book}.

After extensive research spanning decades, two distinct types of progenitors have been identified for GRBs, namely core collapse from Wolf–Rayet stars
for long GRBs \citep{Woosley1993APJ,Paczynski1998,MacFadyen1999APJ,Woosley2006ARAA} and mergers of two compact stellar objects (neutron star–neutron star and neutron star–black hole systems) for short GRBs \citep{Paczynski1986APJL,Eichler1989Nature,Paczynski1991bGRB,Paczynski1991aACTAA,Narayan1992APJL,Abbott2017PhysRevLett}.
Following the catastrophic destruction of the progenitor system, a central engine is thought to form, which powers a relativistic jet. The prompt emission of GRBs is generally believed to originate from the dissipation process of the magnetic energy or kinetic energy of the jet \citep{Rees2005ApJ,Lazzati2009ApJ,Lazzati2013ApJ}, whereas the subsequent afterglow emission is attributed to the interaction between the jet and the circumburst medium \citep{Meszaros1997ApJ}. Therefore, the characteristics of the jet predominantly govern the multi-band radiation properties of GRBs. 

In previous studies, some structured jet models have been proposed, including the power-law jet model \citep{Meszaros1998APJ,Dai2001APJ,Rossi2002MNRASa,Zhang2002APJ,Granot2003APJ} the Gaussian jet model \citep{Zhang2002APJ,Granot2003APJ,Zhang2004APJL} and the two-component jet model \citep{Ramirez-Ruiz2002,Zhang2004,Peng2005ApJ}.
Recently, motivated by the potential of gravitational wave astronomy in relation to GRB sources, the discussions on structured jet become revived \citep[e.g.][]{Lazzati2017MNRAS,Lamb2017MNRAS}. It appears that the jets associated with GRBs probably exhibit a structured nature. This assertion is supported by the results of multi-band observations conducted on short GRB 170817A, which represents the first electromagnetic counterpart of a gravitational wave originating from the merger of binary neutron stars \citep{Abbott2017PhysRevLett,Gao2018SCPMA,Zhang2018NatureCommunications,Gottlieb2018MNRAS,Kasliwal2017Science,Piro2018ApJ,Xiao2017ApJL,Lazzati2018PRL,Lyman2018NatureAstronomy,Troja2018MNRAS}. 
Based on the analysis of GRB 221009A, the most bright GRB ever detected, some studies suggest that the jets of long GRBs may also show structured nature \citep{An2023arXiv,OConnorSci}. A shared characteristic among these jet structures is their symmetric configuration relative to the axis of the jet, and the prompt and afterglow radiation characteristics of GRBs in such models have been extensively analyzed \cite[e.g.][]{Filgas2011A&A,Nicuesa2011A&A,Lamb2017MNRAS,Gill2018MNRAS,Lyman2018NatureAstronomy,Margutti2018ApJL,Resmi2018ApJ,Troja2018MNRAS,Xie2018ApJ,Kann2018A&A,Lamb2019ApJL,Meng2019ApJ,Beniamini2020MNRAS,Oganesyan2020ApJ,Gottlieb2021MNRAS}.

On the other hand, the non-axisymmetric structures have also been studied in the literature. \cite{Meszaros1998APJ} first claimed that, due to the angular anisotropy of the fireball, the afterglow could be significant different with the isotropic scenario. Later, some works investigated the observational features for several possible asymmetric structures, such as the jet hotspots, the patchy shells and the micro/sub jets \citep{Nakamura2000ApJL,Yamazaki2004APJL,Ioka2005ApJ}. 
%The features include short gamma ray emission, X-ray flash, and the variability on the afterglow. 
Recently, \citet{Lamb2022Univ} used the results of the 3-dimensional hydrodynamic jets in the neutron star merger environment to determine the degree of polar and rotational inhomogeneity ($N\times N$ jet model). They found that the result of these inhomgeneities in the jet's energy/Lorentz factor distribution showed some degree of rotational variation, although the change in energy/Lorentz factor from these simulations was not large enough to show significant temporal variability on the afterglow. It is worth noting that in some special cases, GRB jets may be heavily non-axisymmetric. For instance, the presence of significant non-uniformity in the internal magnetic dissipation of a jet can lead to the development of complex and asymmetric jet structures \citep{Narayan2009MNRAS}. A more recent study conducted by \citet{Huang2019MNRAS} has demonstrated that the non-uniformity of jets can exist in the circumferential direction due to the precession of GRBs' central engine. 

Here we intend to conduct a first step analysis of the potential characteristics of gamma-ray burst afterglows within the context of non-axisymmetric structured jets. To achieve this, we will examine a basic jet structure consisting of $N$ partitioned elements around its circumference, where the initial Lorentz factor $\gamma_0$ and isotropic energy $E_{\text{iso}}$ are step functions. This structure can be extended to any arbitrary $N$-value, allowing for the construction of complex asymmetric jet structures. To calculate the afterglow properties for any $N$-value, we have developed a method that utilizes both semi-analytical and numerical estimations. We present an overview of our findings for $N=2$ and $N=4$, and briefly discuss the expected results for any arbitrary $N$-value.

\section{Model description}
\label{model}

\begin{figure}
    \centering
    \includegraphics[width=8.5cm]{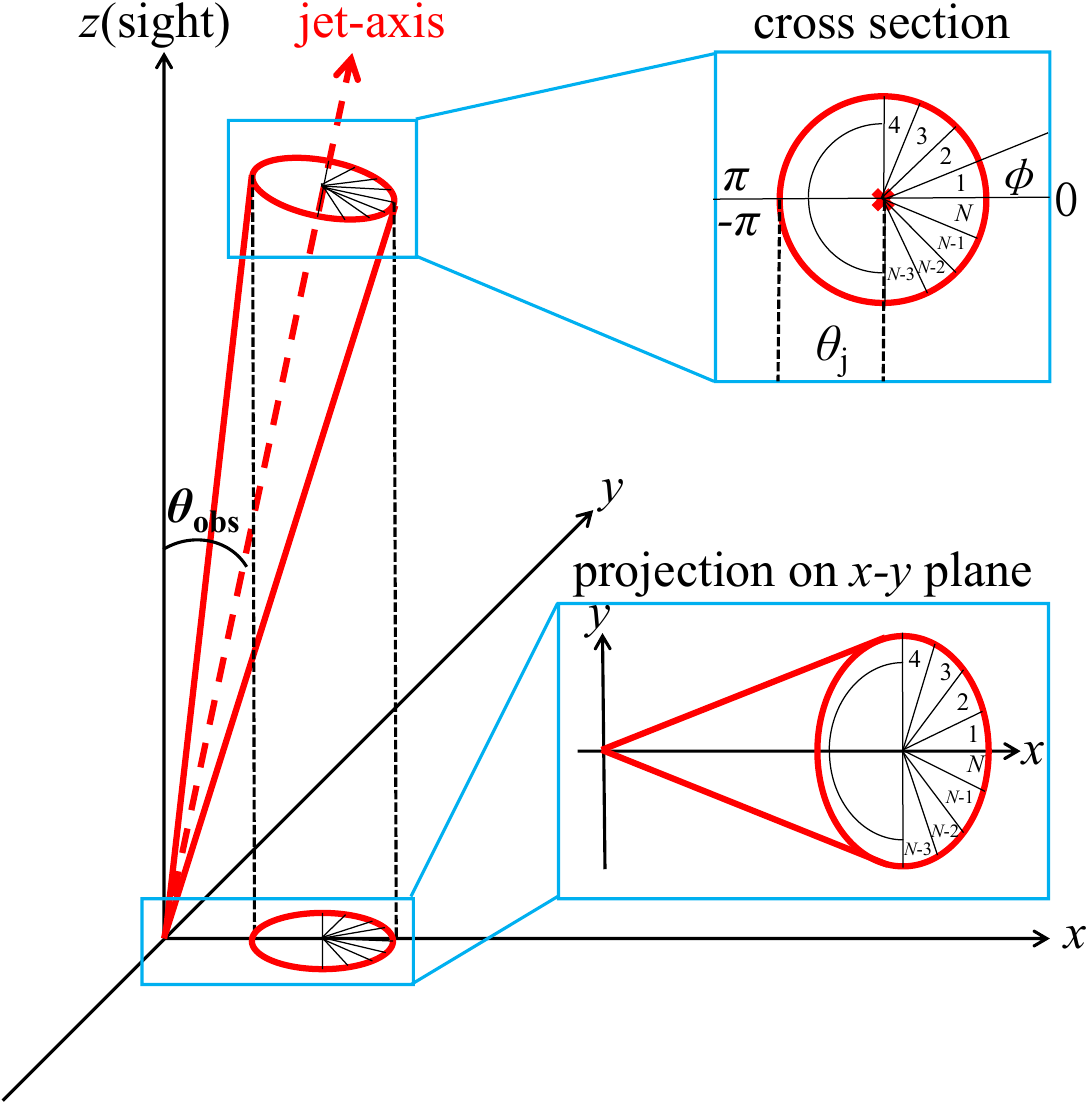}
    \caption{
    The diagram illustrates the jet structure and coordinate system. The example shown on the left side represents an off-axis observation of an asymmetrically structured jet. The upper right section displays a cross-sectional view of the jet, while the lower right section depicts the projection of the jet on the $x-y$ plane. The azimuthal angle $\phi$, ranging from $-\pi$ to $\pi$, is used to denote orientation around the jet axis. This azimuthal range can be divided into $N$ segments to effectively represent the asymmetric nature of the jet structures.  }
    \label{coordinate}
\end{figure}

A non-axisymmetric structured jet can be represented by a schematic image within a coordinate system that combines spherical and Cartesian coordinates. The $z-$axis of this coordinate system points towards the observer, while the jet axis resides in the $x-z$ plane, with the angle between $z-$axis and jet axis as $\theta_{\text{obs}}$ (see Figure \ref{coordinate}). The spherical coordinates are based on the jet axis, with a half-opening angle of a cone around the jet axis denoted as $\theta$, ranging from $\theta=0$ at the jet axis to the half-opening angle of the jet, $\theta_j$. The azimuth angle $\phi$ forms a circumference around the jet axis, ranging from $-\pi$ to $\pi$. We define $\phi=0$ at the projection of the $x$-axis onto the jet's cross-section. The jet is divided into $N$ partitioned elements along the azimuthal direction, while we assume that the jet is uniform along $\theta$. To avoid confusion on the sign of $\theta$ and $\phi$, we only consider the jet in the region $z>0$, so that $ 0 < \theta_{\rm obs} < \pi/2$ and the element at $\phi = 0$ or $\phi = \pi$ is always the furthermost part from the observer's line of sight (LOS).

Here we treat each of the $N$ elements as an independent "patchy", characterized by its own initial Lorentz factor $\gamma_0$ and isotropic kinetic energy $E_{\rm{K, iso}}$. The interaction between each  "patchy" and the interstellar medium could produce a strong external shock. Electrons are accelerated in the external shocks, which radiate synchrotron emission in the magnetic fields behind the shocks that are believed to be generated in situ due to plasma instabilities \citep[][for a review]{Gao2013NewAR}.

In most cases (except for $\theta_{\rm obs}=0$), the majority of "patchies" are off-axis with respect to the observer. For the $i$-th "patchy", we can first use the standard GRB afterglow model to calculate its on-axis flux evolution with time, $F_{\nu,i}(t)$ (see sections \ref{sec:numerical_formalism} and \ref{sec:maths} for details), then we can transfer $F_{\nu,i}(t)$ to the observer direction through Doppler conversion. The cumulative effect of the individual contributions from $N$ 'patchies' can yield the comprehensive afterglow characteristics of a non-axisymmetric jet. 

\subsection{Numerical Formalism}
\label{sec:numerical_formalism}
For the $i$-th "patchy" (henceforth treated as a uniform jet), we can first follow the formulae derived in \citet{Huang2000ApJ} to calculate its dynamical evolution. 

In the frame of an on-axis observer, the evolution of the jet's radius $R$ over time $T$ reads as
\begin{equation}
    \frac{dR}{dT}=\beta c\gamma(\gamma +\sqrt{\gamma^2-1}),
    \label{eq:dR_dT}
\end{equation}
where $\beta$ and $\gamma$ represent the dimensionless velocity and Lorentz factor of the jet's bulk motion, respectively. The accumulation of the jet-swept mass by each element from the interstellar medium $m$ with the jet radius $R$ can be described as
\begin{equation}
    \frac{dm}{dR}=\frac{1}{N}2\pi R^2(1-\cos{\theta_{\text{j}}})nm_{\text{p}},
\end{equation}
where $m_{\text{p}}$ is the mass of proton. $n=AR^{-k}$ is the particle number density of the interstellar medium, where $k$ is wind profile variable. $k=0$ is for uniform interstellar medium and $k=2$ is for stellar wind environment. For the wind model, $A=(3.0\times10^{35}{\rm cm^{-1}})A_*$, $A_*$ is a dimensionless free parameter depending on the wind environment \citep{2018pgrb.book}. $\theta_{\text{j}}$ stands for the half-opening angle of the jet. 
Here we ignore the lateral spread of the jet, so that $\theta_{\text{j}}$ is treated as a constant.  
Taking the radiation cooling effect into consideration, the evolution of jet's bulk motion Lorentz factor $\gamma$ with respect to $m$ can be written as
\begin{equation}
   \frac{d\gamma}{dm}=-\frac{\gamma^2-1}{M_{\text{ej}}+\varepsilon m+2(1-\varepsilon)\gamma m},
   \label{eq:dgammma_dm}
\end{equation} 
where $M_{\text{ej}}=E_0/\left(\gamma_0c^2\right)$ is the ejecta mass and $E_0$ is the initial kinetic energy of a element. The radiative efficiency, $\varepsilon$, is defined as the fraction of the shock generated internal energy (in jet's comoving frame) that would be radiated, which can be expressed as \citep{Dai1999ApJL} 
\begin{equation}  \varepsilon=\epsilon_e\frac{t_{\text{syn}}^{'-1}}{t_{\text{syn}}^{'-1}+t_{\text{ex}}^{'-1}},
\end{equation}
where $t^{'}_{\text{syn}}=6\pi m_ec/\left(\sigma_{\text{T}}B^{'2}\gamma_{e,\text{min}}\right)$ is the synchrotron cooling timescale, and $t^{'}_{\text{ex}}=R/\left(\gamma c\right)$ is the expansion timescale in the jet's comoving frame. $m_e$ represents the mass of electron, and $\sigma_{\text{T}}$ represents the cross section for Thompson scattering. Assume a fraction $\epsilon_B$ of the total shock-generated internal energy goes into the random magnetic field, the magnetic energy density in the jet's comoving frame can thus be estimated as
\begin{equation}
    \frac{B^{'2}}{8\pi}=\epsilon_B^2\frac{\hat{\gamma}\gamma+1}{\hat{\gamma}-1}\left(\gamma-1\right)nm_pc^2, 
\end{equation}
where $\hat{\gamma}=\left(4\gamma+1\right)/\left(3\gamma\right)$ is the adiabatic index \citep{Dai1999ApJL}. 
Assume a fraction $\epsilon_e$ of the total shock-generated internal energy goes into the electrons and assume that the accelerated electrons is a power law
function with the index of $p$ ($dN_e/d\gamma_e \propto \gamma_e^{-p}$), the minimum Lorentz factor for the random motion of electrons in the jet's comoving frame can thus be derived as \citep{Huang2000ApJ}
\begin{equation}
\gamma_{e,\text{min}}=\epsilon_e\left(\gamma-1\right)\frac{m_p\left(p-2\right)}{m_e\left(p-1\right)}+1.
\end{equation} 

For synchrotron radiation, the observed radiation power and the characteristic frequency of an electron with Lorentz factor $\gamma_e$ are given by \citep{Sari1998ApJL}
\begin{equation}   P\left(\gamma_e\right)=\frac{4}{3}\sigma_Td\gamma^2\gamma_e^2\frac{B^2}{8\pi}.
   \label{eq:P_gammae}
\end{equation}
\begin{equation}   \nu\left(\gamma_e\right)=\gamma\gamma_e^2\frac{q_eB}{2\pi m_ec},
   \label{eq:nu_char}
\end{equation}
where $q_e$ is the charge of an electron. The peak power occurs at  $\nu(\gamma_e)$, where it has the approximate value
\begin{equation}
    P_{\nu,\text{max}}\approx\frac{P\left(\gamma_e\right)}{\nu\left(\gamma_e\right)}=\frac{m_ec^2\sigma_T}{3q_e}\gamma B.
\end{equation}
Usually, a characteristic Lorentz factor $\gamma_c$ is defined as \citep{Sari1998ApJL}
\begin{equation}
    \gamma_c=\frac{6\pi m_ec}{\sigma_TB^2T}=\frac{3m_e}{16\epsilon_B\sigma_Tm_pc}\frac{1}{T\gamma^3n},
\end{equation}
beyond which the electrons might have significantly cooled. 

The electrons' Lorentz factors $\gamma_{e,\text{min}}$ and $\gamma_c$ define two characteristic emission frequencies $\nu_m$ and $\nu_c$ in the synchrotron spectrum. For the fast cooling regime ($\nu_c<\nu_m$), the self absorption frequency $\nu_a$ is
\begin{equation}
\nu_a=\begin{cases}
\left[\frac{c_1q_enR}{\left(3-k\right)B\gamma_c^5}\right]^{3/5}\nu_c &\nu_a<\nu_c,\\
\left[\frac{c_2q_enR}{\left(3-k\right)B\gamma_c^5}\right]^{1/3}\nu_c &\nu_c<\nu_a<\nu_m,\\
\left[\frac{c_2q_enR}{\left(3-k\right)B\gamma_c^5}\right]^{2/\left(p+5\right)}\left(\frac{\nu_m}{\nu_c}\right)^{\left(p-1\right)/\left(p+5\right)}\nu_c &\nu_m<\nu_a.
\end{cases}
\end{equation}
$c_1$ and $c_2$ are coefficients dependent on $p$ \citep{Wu2003MNRAS}. The observed flux density $F_{\nu}$ is divided into the following three situations\\
(1)$\nu_a<\nu_c<\nu_m$: 
\begin{equation}
\label{fc1}
F_{\nu}=F_{\nu,\text{max}}\begin{cases}
\left(\frac{\nu}{\nu_a}\right)^2\left(\frac{\nu_a}{\nu_c}\right)^{1/3} &\nu<\nu_a,\\
\left(\frac{\nu}{\nu_c}\right)^{1/3} &\nu_a<\nu<\nu_c,\\
\left(\frac{\nu}{\nu_c}\right)^{-1/2} &\nu_c<\nu<\nu_m,\\
\left(\frac{\nu_m}{\nu_c}\right)^{-1/2}\left(\frac{\nu}{\nu_m}\right)^{-p/2} &\nu_m<\nu.
\end{cases}
\end{equation}
(2)$\nu_c<\nu_a<\nu_m$: 
\begin{equation}
\label{fc2}
F_{\nu}=F_{\nu,\text{max}}\begin{cases}
\left(\frac{\nu}{\nu_c}\right)^2\left(\frac{\nu_c}{\nu_a}\right)^3 &\nu<\nu_c,\\
\left(\frac{\nu}{\nu_a}\right)^{5/2}\left(\frac{\nu_a}{\nu_c}\right)^{-1/2} &\nu_c<\nu<\nu_a,\\
\left(\frac{\nu}{\nu_c}\right)^{-1/2} &\nu_a<\nu<\nu_m,\\
\left(\frac{\nu_m}{\nu_c}\right)^{-1/2}\left(\frac{\nu}{\nu_m}\right)^{-p/2} &\nu_m<\nu.
\end{cases}
\end{equation}
(3)$\nu_c<\nu_m<\nu_a$: 
\begin{equation}
\label{fc3}
F_{\nu}=F_{\nu,\text{max}}\begin{cases}
\left(\frac{\nu}{\nu_c}\right)^2\left(\frac{\nu_c}{\nu_a}\right)^3\left(\frac{\nu_a}{\nu_m}\right)^{-\left(p-1\right)/2} &\nu<\nu_c,\\
\left(\frac{\nu}{\nu_a}\right)^{5/2}\left(\frac{\nu_a}{\nu_m}\right)^{-p/2}\left(\frac{\nu_m}{\nu_c}\right)^{-1/2} &\nu_c<\nu<\nu_a,\\
\left(\frac{\nu}{\nu_m}\right)^{-p/2}\left(\frac{\nu_m}{\nu_c}\right)^{-1/2} &\nu_a<\nu.
\end{cases}
\end{equation}
where $F_{\nu,\text{max}}$ represents the peak flux density, which can be estimated as 
\begin{equation}
    F_{\nu,\text{max}}=\frac{N_eP_{\nu,\text{max}}}{4\pi D_L^2},
\end{equation}
where $N_e$ is the total number of swept-up electrons in the post-shock fluid (assuming a spherical geometry). $D_L$ is the luminosity distance from the source to the observer. And in the slow cooling regime($\nu_c>\nu_m$), the self absorption frequency $\nu_a$ is:
\begin{equation}
\nu_a=F_{\nu,\text{max}}\begin{cases}
\left[\frac{c_1q_enR}{\left(3-k\right)B\gamma_c^5}\right]^{3/5}\nu_m &\nu_a<\nu_m,\\
\left[\frac{c_2q_enR}{\left(3-k\right)B\gamma_c^5}\right]^{2/\left(p+4\right)}\nu_m &\nu_m<\nu_a<\nu_c,\\
\left[\frac{c_2q_enR}{\left(3-k\right)B\gamma_c^5}\right]^{2/\left(p+5\right)}\left(\frac{\nu_c}{\nu_m}\right)^{1/\left(p+5\right)}\nu_m &\nu_c<\nu_a.
\end{cases}
\end{equation}
And the flux in slow cooling regime is\\
(1)$\nu_a<\nu_m<\nu_c$: 
\begin{equation}
\label{lc1}
F_{\nu}=F_{\nu,\text{max}}\begin{cases}
\left(\frac{\nu}{\nu_a}\right)^2\left(\frac{\nu_a}{\nu_m}\right)^{1/3} &\nu<\nu_a,\\
\left(\frac{\nu}{\nu_m}\right)^{1/3} &\nu_a<\nu<\nu_m,\\
\left(\frac{\nu}{\nu_m}\right)^{-\left(p-1\right)/2}F_{\nu,\text{max}} &\nu_m<\nu<\nu_c,\\
\left(\frac{\nu}{\nu_m}\right)^{-\left(p-1\right)/2}\left(\frac{\nu}{\nu_c}\right)^{-p/2} &\nu_c<\nu.
\end{cases}
\end{equation}
(2)$\nu_m<\nu_a<\nu_c$: 
\begin{equation}
\label{lc2}
F_{\nu}=F_{\nu,\text{max}}\begin{cases}
\left(\frac{\nu}{\nu_m}\right)^2\left(\frac{\nu_m}{\nu_a}\right)^{\left(p+4\right)/2}F_{\nu,\text{max}} &\nu<\nu_m,\\
\left(\frac{\nu}{\nu_a}\right)^{5/2}\left(\frac{\nu_a}{\nu_m}\right)^{-\left(p-1\right)/2}F_{\nu,\text{max}} &\nu_m<\nu<\nu_a,\\
\left(\frac{\nu}{\nu_m}\right)^{-\left(p-1\right)/2}F_{\nu,\text{max}} &\nu_a<\nu<\nu_c,\\
\left(\frac{\nu}{\nu_m}\right)^{-\left(p-1\right)/2}\left(\frac{\nu}{\nu_c}\right)^{-p/2}F_{\nu,\text{max}} &\nu_c<\nu.
\end{cases}
\end{equation}
(3)$\nu_m<\nu_c<\nu_a$: 
\begin{equation}
\label{lc3}
F_{\nu}=F_{\nu,\text{max}}\begin{cases}
\left(\frac{\nu}{\nu_m}\right)^2\left(\frac{\nu_m}{\nu_a}\right)^{\left(p+4\right)/2}\left(\frac{\nu_a}{\nu_c}\right)^{-1/2} &\nu<\nu_m,\\
\left(\frac{\nu}{\nu_a}\right)^{5/2}\left(\frac{\nu_a}{\nu_c}\right)^{-p/2}\left(\frac{\nu_c}{\nu_m}\right)^{-\left(p-1\right)/2} &\nu_m<\nu<\nu_a,\\
\left(\frac{\nu}{\nu_c}\right)^{-p/2}\left(\frac{\nu_c}{\nu_m}\right)^{-\left(p-1\right)/2} &\nu_a<\nu.
\end{cases}
\end{equation}

For an off-axis observer, the observed flux needs to be corrected by\footnote{For more precise results, it is better to calculate the flux for a given observation angle throughout the flux calculation \citep[e.g,][]{Lamb2018MNRAS,Fraija2020ApJ,Ryan2020ApJ,Nedora2023MNRAS}.} \citep{Granot2002ApJL} 
\begin{equation}
F_{\nu}=a^3F_{\nu/a}\left(at\right),
   \label{lzf}
\end{equation}
with a factor
\begin{equation}
    a=\frac{1-\beta}{1-\beta\cos{\theta_{\rm obs}}}\approx\frac{1}{1+\gamma^2\theta_{\rm obs}^2},
    \label{lz}
\end{equation}
where $\theta_{\rm obs}$ is the angle between LOS and the jet. In this work, if the LOS pass through the $i-$th element of the jet, we take $\theta_{\rm obs,i}=0$, otherwise, we take the angle between the LOS and the nearest edge of the $i-$th element as $\theta_{\rm obs,i}$. Overall, the total radiation flux can be calculated as 
\begin{equation}
F_{\nu}\left(t\right)=\sum\limits^N\limits_{i=1}a_{i}^3F_{\nu/a_{i}}\left(a_{i}t\right). 
\label{sum}
\end{equation}

\subsection{Semi-analytical Formalism}
\label{sec:maths} % used for referring to this section from elsewhere
In addition to the numerical approach, \cite{Granot2005ApJ} presented a semi-analytical technique for characterizing the afterglow of GRBs with a structured jet, although their work only considered the jet's polar angle ($\theta$) dependence. Considering that semi-analytical results can help us better understand the properties of light curve results, such as the peak time and the rising and decaying slopes, we have also provided a semi-analytical formalism for the non-axisymmetric jet model. 
To model the afterglow of a relativistic jet structured with azimuthal variation, akin to the methodology described in Section \ref{sec:numerical_formalism}, we divide the entire jet into $N$ segments along the azimuthal angle and analyze them independently using the methodology outlined in \citet{Granot2005ApJ}. 

For each independent element, 
the evolution of the bulk-motion Lorentz factor $\gamma$ can be approximately expressed as a function of ejecta's radius $R$, which reads as 
\citep{Blandford1976PhFl}:
\begin{equation}
    \gamma\left(R\right)\approx\begin{cases}
    \gamma_0&R<R_{\text{dec}},\\
    \gamma_0\left(R/R_{\text{dec}}\right)^{-\left(3-k\right)/2}&R>R_{\text{dec}},
    \end{cases}
    \label{gr}
\end{equation}
where $\gamma_0$ is the initial bulk-motion Lorentz factor and $R_{\text{dec}}$ is the deceleration radius. Before the deceleration time ($T < T_{\rm dec}$), the observed flux (for an off-axis observer) can be calculated as
\begin{equation}
    \label{F1}
    \begin{split}
        F_{\nu}\left(T\right)&=\frac{2\gamma_0L'_{\nu/2\gamma_0}\left[R_L\left(T\right)\right]}{4\pi D_L^2}\int_0^1dxx^{1+\alpha-\beta}\frac{\Delta\phi\left(x\right)}{2\pi}\\
        &=\frac{2\gamma_0L'_{\nu/2\gamma_0}\left(R_{\text{dec}}\right)}{4\pi D_L^2}\left(\frac{T}{T_{\text{dec}}}\right)^\alpha\int_0^1dxx^{1+\alpha-\beta}\frac{\Delta\phi\left(x\right)}{2\pi}.
    \end{split}
\end{equation}
while after that ($T > T_{\rm dec}$) the flux reads as
\begin{equation}
    \label{F2}
    \begin{split}
        &F_{\nu}\left(T\right)=\frac{2\gamma_0L'_{\nu/2\gamma_0}\left(R_{\text{dec}}\right)}{4\pi D_L^2}\left\{\left(\frac{T}{T_{\text{dec}}}\right)^{\beta-2} \right .\\
        &\times\int^1_0dyy^{1+\alpha-\beta}\frac{\Delta\phi\left(y\right)}{2\pi}+x_{\text{dec}}^{-\alpha+\left(1-\beta\right)\left(3-k\right)/2}\\
        &\left .\times\int_{x_{\text{dec}}}^1dxx^{\alpha-2+\left(3-\beta\right)\left(5-k\right)/2}\left[\frac{1+3\left(3-k\right)x^{4-k}}{4-k}\right]^{\beta-2}\frac{\Delta\phi\left(x\right)}{2\pi}\right\},
    \end{split}
\end{equation} 
where the power-law indices $\alpha$ and $\beta$ change between different power-law segments (PLSs) of the spectrum, which are listed in Table \ref{PLS} \citep{Granot2002ApJ}. The integral variables are defined as $x=R/R_L$ and $y = R/R_{\rm dec}$, thus $x_{\rm dec} = R_{\rm dec}/R_L$.
$\gamma_L$ and $R_L\left(T\right)$ are the Lorentz factor and radius when a photon is emitted and reaches the observer at time $T$ in the observer's frame. We have
\begin{equation}
    R_L\left(T\right)=\frac{2cT}{1+z}\begin{cases}
    \gamma_0^2&T\leq T_{\text{dec}},\\
    \frac{\left(4-k\right)\gamma_L^2}{1+\left(4-k\right)x_{\text{dec}}^{4-k}}&T>T_{\text{dec}}.
    \end{cases}
\end{equation}
where $z$ is the red shift of the source.
\begin{table}
	\caption{Indexes $a$ and $b$ at different power-law segments (PLSs) of the spectrum. $p$ represents the power-law index for the electron energy distribution.}
	\label{PLS}
	\begin{tabular}{cccc} % four columns, alignment for each
		\hline
		PLS & $\beta$ & $\alpha$($R<R_{\text{dec}}$) & $\alpha$($R>R_{\text{dec}}$)\\
		\hline
		D & $1/3$ & $3-k/2$ & $3-4k/3$\\
		E & $1/3$ & $11/3-2k$ & $\left(5-4k\right)/3$\\
		F & $-1/2$ & $2-3k/4$ & $\left(5-2k\right)/4$\\
		G & $\left(1-p\right)/2$ & $3-k\left(p+5\right)/4$ & $\left[15-9p-2k\left(3-p\right)\right]/4$\\
		H & $-p/2$ & $2-k\left(p+2\right)/4$ & $\left[14-9p+2k\left(p-2\right)\right]/4$\\
		\hline
	\end{tabular}
	\label{PLS}
\end{table}
$L^{\prime}_{\nu^{\prime}}$ represents the specific luminosity of the afterglow in the jet's comoving frame, where $\nu^{\prime} \approx \nu / 2\gamma_0$, while $\nu$ is defined in the observer's frame. The coefficient in front of the time term in Equation \ref{F1} and \ref{F2} is approximately equal to the flux density at the deceleration time $F_{\nu}\left(T_{\text{dec}}\right)$, where the deceleration time can be calculate by
\begin{equation}
T_{\text{dec}}=\begin{cases}
    90.5\left(1+z\right)n^{-1/3}E_{\text{iso},52}^{1/3}\gamma_{0,2}^{-8/3}s&k=0,\\
    0.3\left(1+z\right)A_{*}^{-1}E_{\text{iso},52}\gamma_{0,2}^{-4}s&k=2,
\end{cases}
\end{equation}

After performing individual computation of the on-axis radiation flux for each element, the summation of radiation flux emanating from all elements in a specified LOS direction can be ascertained through the utilization of formulas \ref{lzf} to \ref{sum}, analogous to the numerical approach.

\section{Non-axisymmetric jet afterglow}
\label{result}
With the formula introduced in Section \ref{model}, here we calculate the light curves of the afterglows in some specific cases. In this section, we only consider uniform interstellar media ($k=0$).
\subsection{Two-element jet with interface in the plane containing the LOS}
\label{subsec:two-element-alinged}
\begin{figure}
    \centering
	\subfigure{\includegraphics[width=8.5cm]{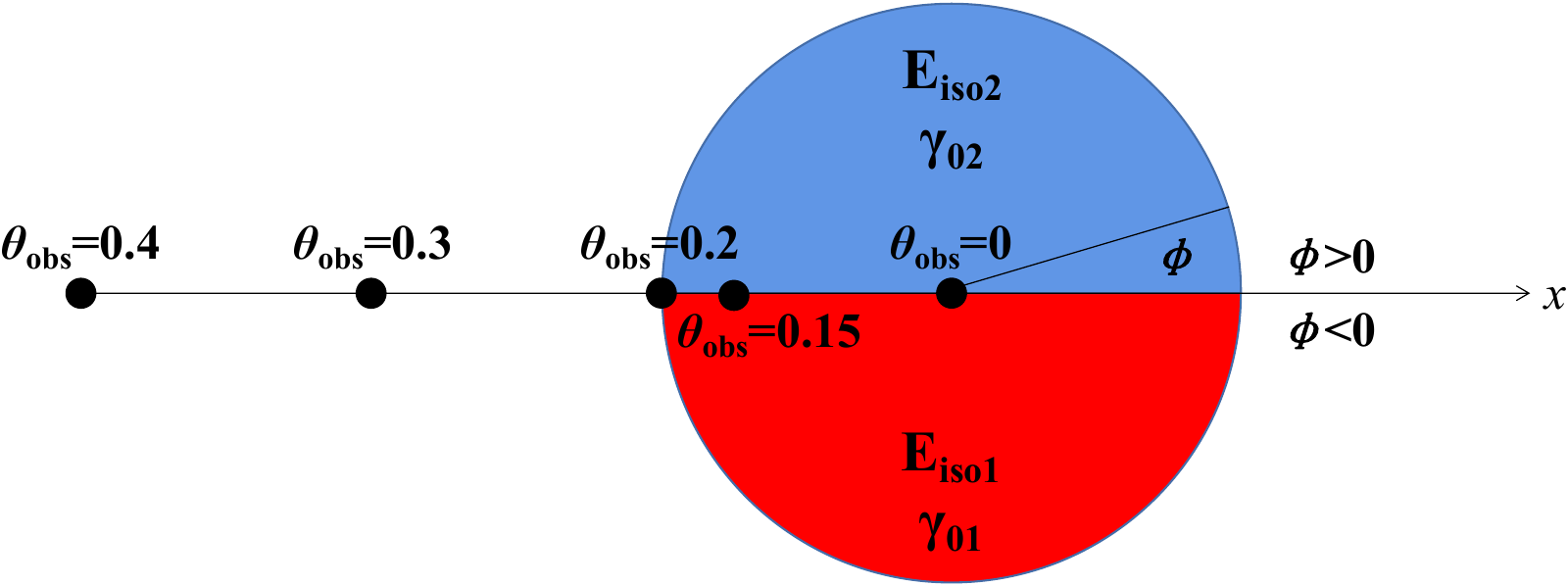}
		\label{3.1.1}}
	\caption{The schematic diagrams depict the cross-sectional view of a two-element jet with a half-open angle $\theta_{\text{j}}=0.2$. The positions of the LOS, represented by $\theta_{\text{obs}}$, are indicated on the $x$-axis to illustrate the relative arrangements.}
	\label{3.1}
\end{figure}

The simplest non-axisymmetric jet structure is characterized by a sharp interface between two distinct elements resulting from variations of the physical parameters $\gamma_0$ and $E_{\text{iso}}$ at different azimuth $\phi$. Assume it is uniform in $\theta$ direction and has a well-defined interface. In order to explore the effects of the physical parameters of two elements on the light variation curve more clearly, we first considered a special case that the plane of interface contains the LOS. On the $x-y$ plane, the projection of the interface is along the $x$-axis. 
This structure can be mathematically described as
\begin{equation}
    \gamma_0=\begin{cases}
    \gamma_{01}&-\pi<\phi<0,\\
    \gamma_{02}&\text{others},
    \end{cases}
    \label{g0}
\end{equation}
and
\begin{equation}
    E_{\text{iso}}=\begin{cases}
    E_{\text{iso},1}&-\pi<\phi<0,\\
    E_{\text{iso},2}&\text{others}.
    \end{cases}
    \label{E0}
\end{equation}
Figure \ref{3.1} shows the cross section for the two-element jet discussed in this paper.

\begin{figure*}
	\centering
	\subfigure[Numerical approach]{\includegraphics[width=8.5cm]{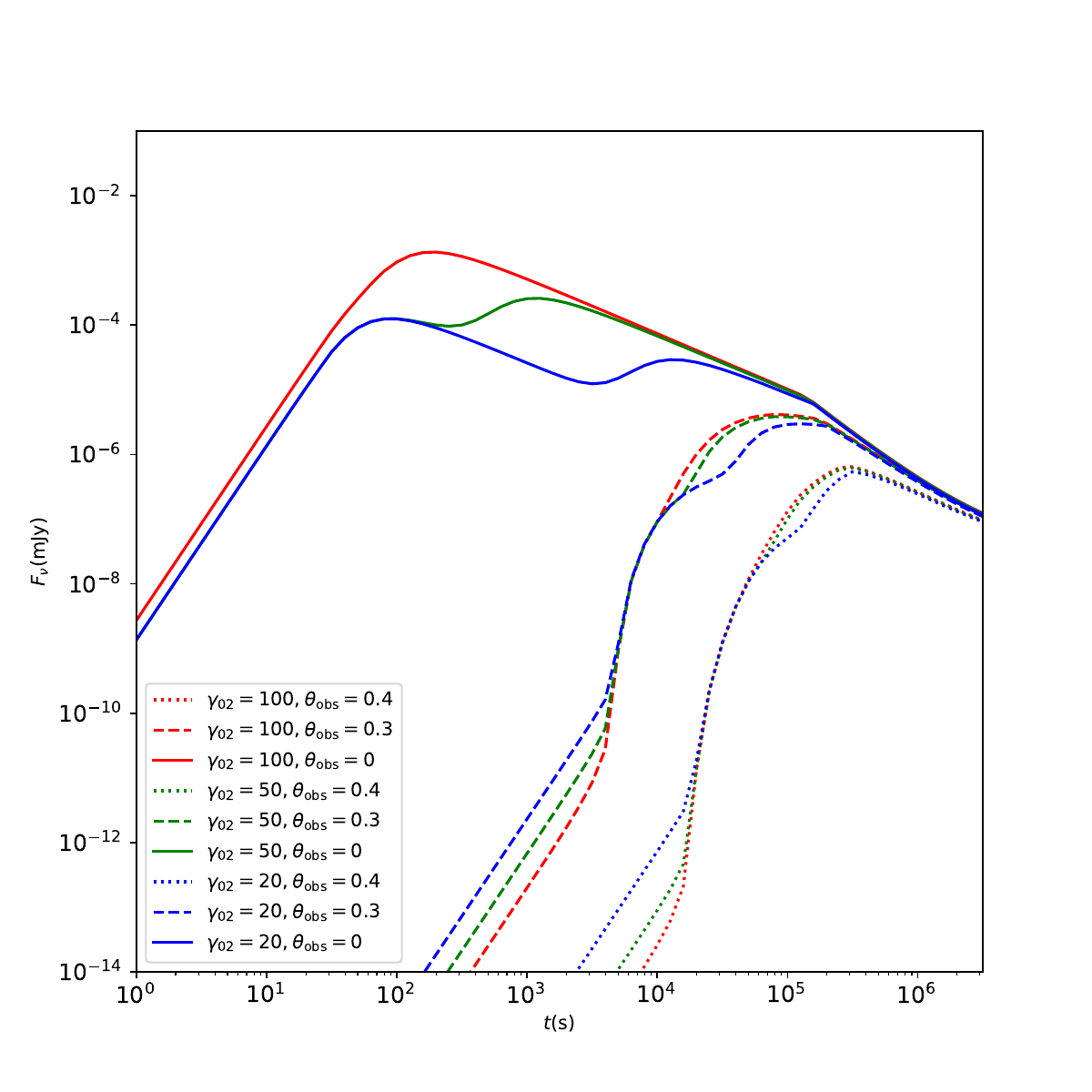}
		\label{szgamma}}
	\subfigure[Semi-analytical approach]{\includegraphics[width=8.5cm]{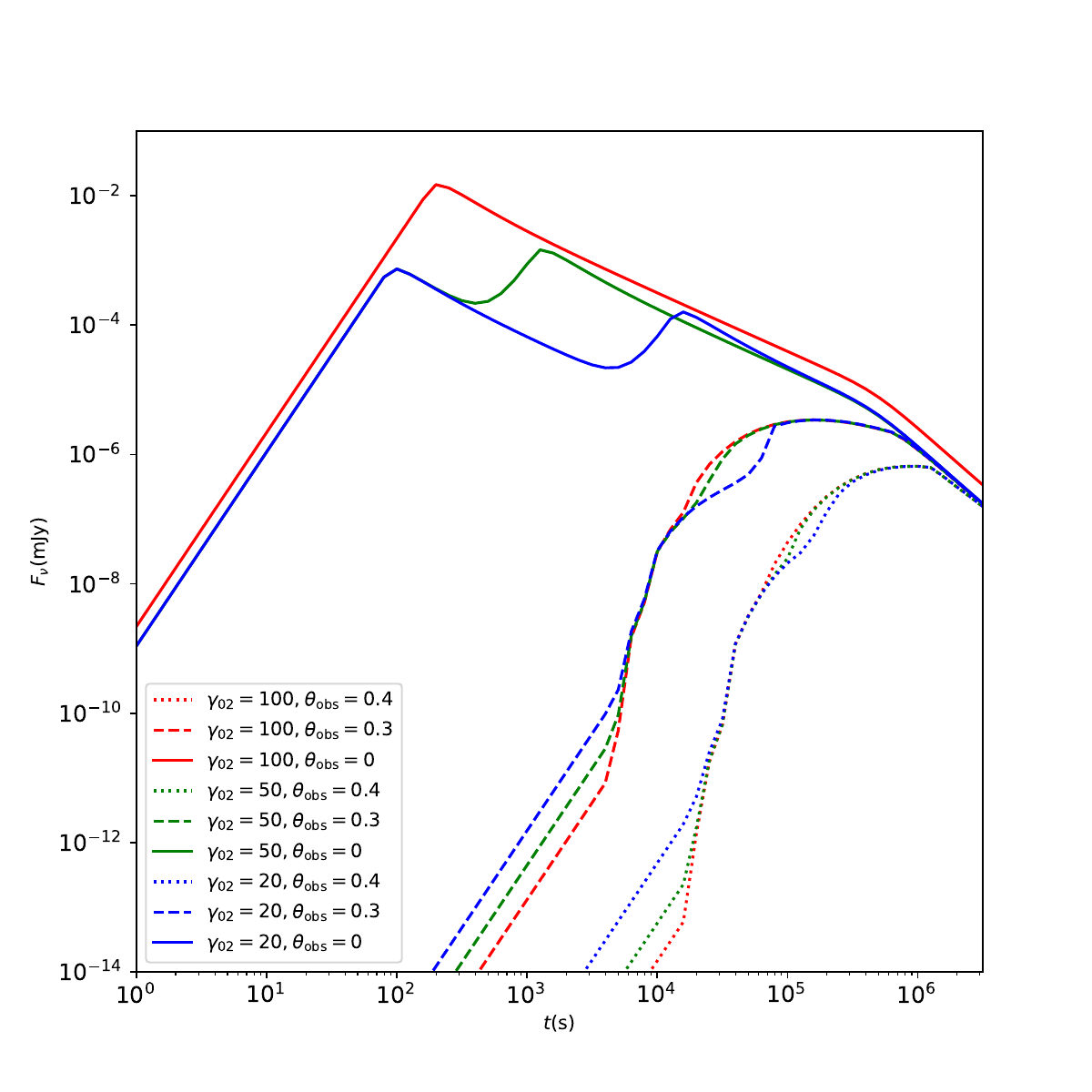}
		\label{jxgamma}}
	\caption{The afterglow's light curves for a two-component jet with the bulk-motion Lorentz factor for one element being fixed at $\gamma_{01}=100$, while the other is varying from $20$ to $100$. The different color represent different value of $\gamma_{02}$. And we use the solid line and dash line to distinct the on-axis observation and off-axis observation.}
	\label{gamma}
\end{figure*}

The light curves with different bulk-motion Lorentz factors for the jet are shown in Figure \ref{gamma}. Specifically, we fix the bulk-motion Lorentz factor for one element at $\gamma_{01}=100$, while systematically increasing the value of the other element from $20$ to $100$. In the plot, we set $E_{\text{iso}1}=10^{50}\text{ergs}$ and $E_{\text{iso}2}=10^{51}\text{ergs}$, a jet's half opening angle of $\theta_{\text{j}}=0.2$, electron power-law distribution spectral index of $p=2.2$, interstellar medium particle number density of $n=0.1 \text{cm}^{-3}$, and microphysics shock parameters, i.e., the electron and magnetic energy fraction parameters $\epsilon_e=0.1$ and $\epsilon_B=0.001$. As an example, we consider an observation frequency of $\nu_{\rm obs}=8.22\times10^{14}$ Hz. The results indicate that the asymmetry of the Lorentz factors in the jet can significantly affect the shape of the afterglow light curve. For an on-axis observer, when the asymmetry reaches a certain level, the light curve exhibits two distinct peaks. As the asymmetry becomes stronger, the time interval between the two peaks gradually increases. For an off-axis observer, the asymmetry of the Lorentz factors in the jet usually results in wiggling during the rising phase, without exhibiting a clear double-peak structure. 

\begin{figure*}
	\centering
	\subfigure[Numerical approach]{\includegraphics[width=8.5cm]{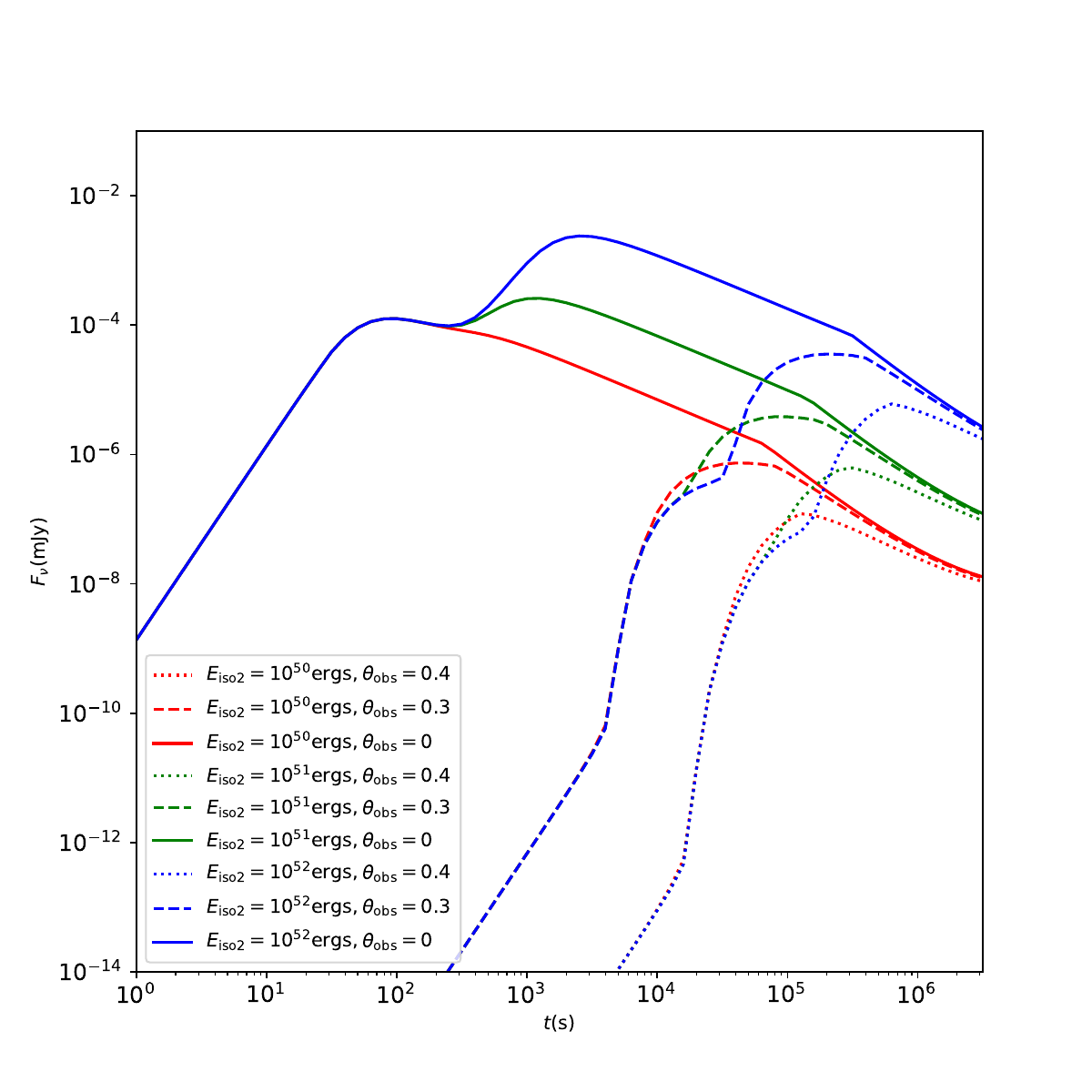}
		\label{szE}}
	\subfigure[Semi-analytical approach]{\includegraphics[width=8.5cm]{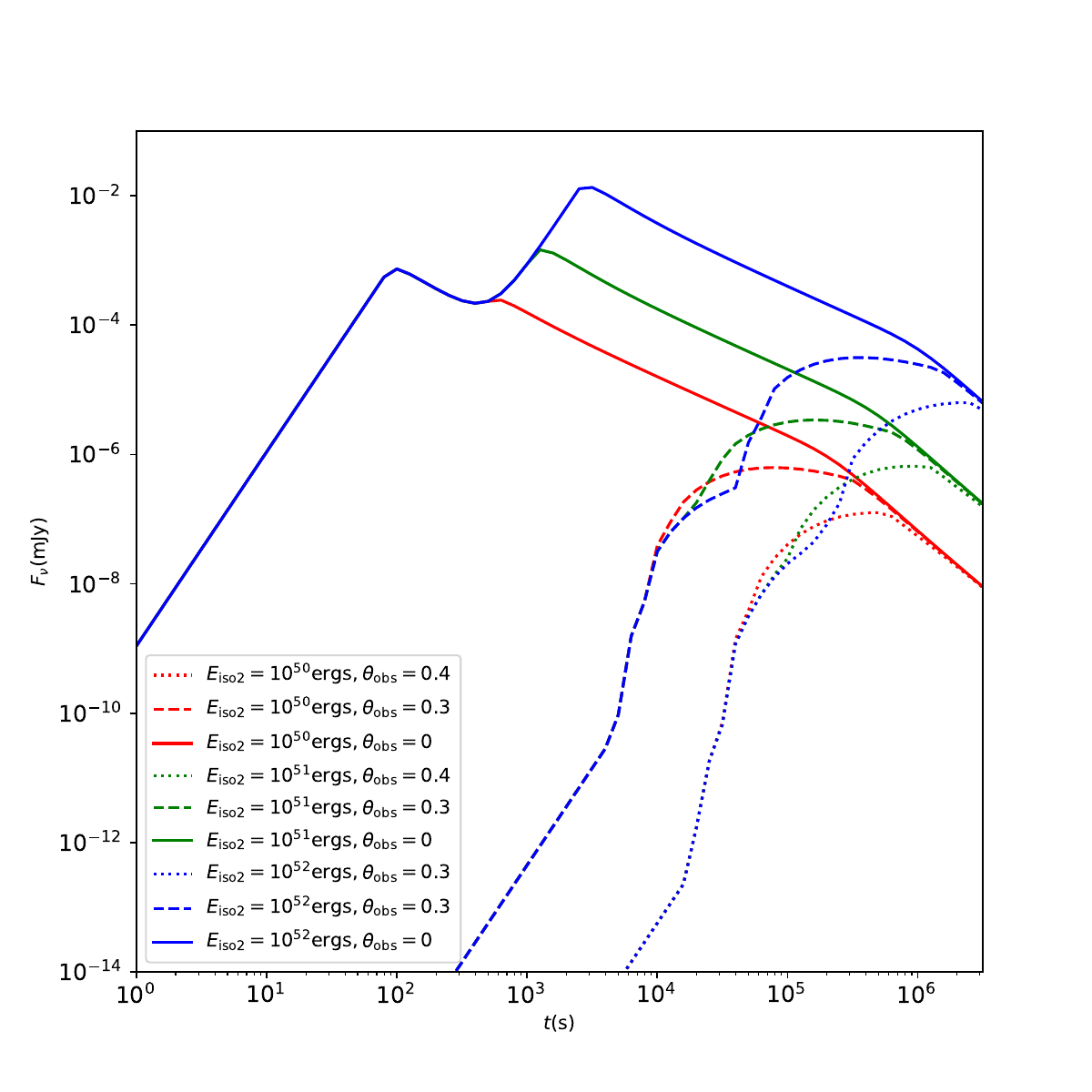}
		\label{jxE}}
	\caption{ 
The afterglow's light curves for a two-element jet with theisotropic energy for one element being fixed at $E_{\text{iso},1}=10^{50} \text{ergs}$ and the other is varying from $10^{50} \text{ergs}$ to $10^{52} \text{ergs}$. The different color represent different value of $E_{\text{iso},2}$. And we use the solid line and dash line to distinct the on-axis observation and off-axis observation.}
	\label{E}
\end{figure*}

Figure \ref{E} illustrates the impact of the isotropic energy $E_{\text{iso}}$ for each element on the afterglow's light curve. Similarly, we fixed the $E_{\text{iso},1}=10^{50} \text{ergs}$ for one element and vary the other from $10^{50} \text{ergs}$ to $10^{52} \text{ergs}$. And we fix the initial Lorentz factor of two elements at $\gamma_{01}=100$ and $\gamma_{02}=50$, with all other parameters identical to those in Figure \ref{gamma}. The results indicate that the asymmetry of the isotropic energy in the jet can also alter the shape of the afterglow light curve. For a given asymmetry in the Lorentz factor of the jet, the larger the energy of the slower portion, the later and brighter the second peak appears in the light curve. For an off-axis observer, although the double-peak structure disappears, the wiggling of the rising segment of the light curve becomes more pronounced with increasing energy of the slower element. 

\subsection{Two-element jet whose interface in the plane intersecting with the LOS}

\begin{figure}
    \subfigure{\includegraphics[width=8.5cm]{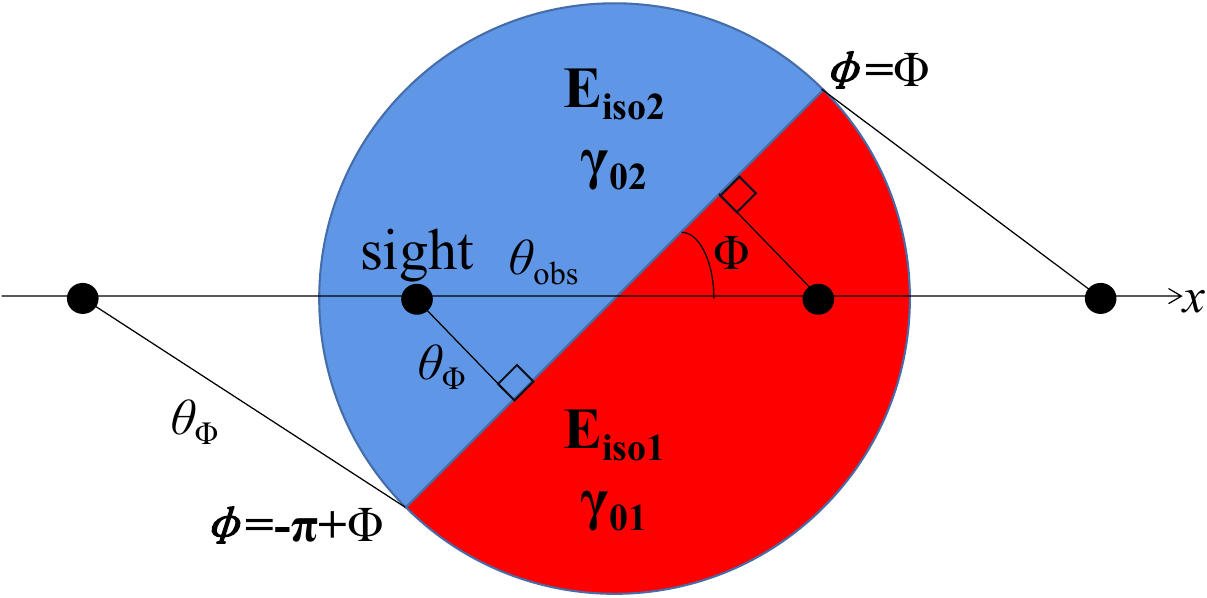}
		\label{3.2.1}}
	\caption{The diagrams of a two-element jet with interface at an arbitrary $\phi$.}
	\label{3.2}
\end{figure}

In most cases, the interface plane between the two elements of the jet will not contain the LOS. In such cases, we denote the azimuth of the interface on the jet's spherical coordinate system relative to the LOS as $\Phi$ (see Figure \ref{3.2}). Equations \ref{g0} and \ref{E0} could be generalized as 

\begin{equation}
    \gamma_0=\begin{cases}
    \gamma_{01}&-\pi+\Phi<\phi<\Phi,\\
    \gamma_{02}&\text{others},
    \end{cases}
    \label{g0r}
\end{equation}
and
\begin{equation}
    E_{\text{iso}}=\begin{cases}
    E_{\text{iso}1}&-\pi+\Phi<\phi<\Phi,\\
    E_{\text{iso}2}&\text{others}.
    \end{cases}
    \label{E0r}
\end{equation}
Figure \ref{3.2} shows the schematic picture for the cases when $\Phi\neq0$.

\begin{figure*}
\centering
	\subfigure[Numerical approach]{\includegraphics[width=8.5cm]{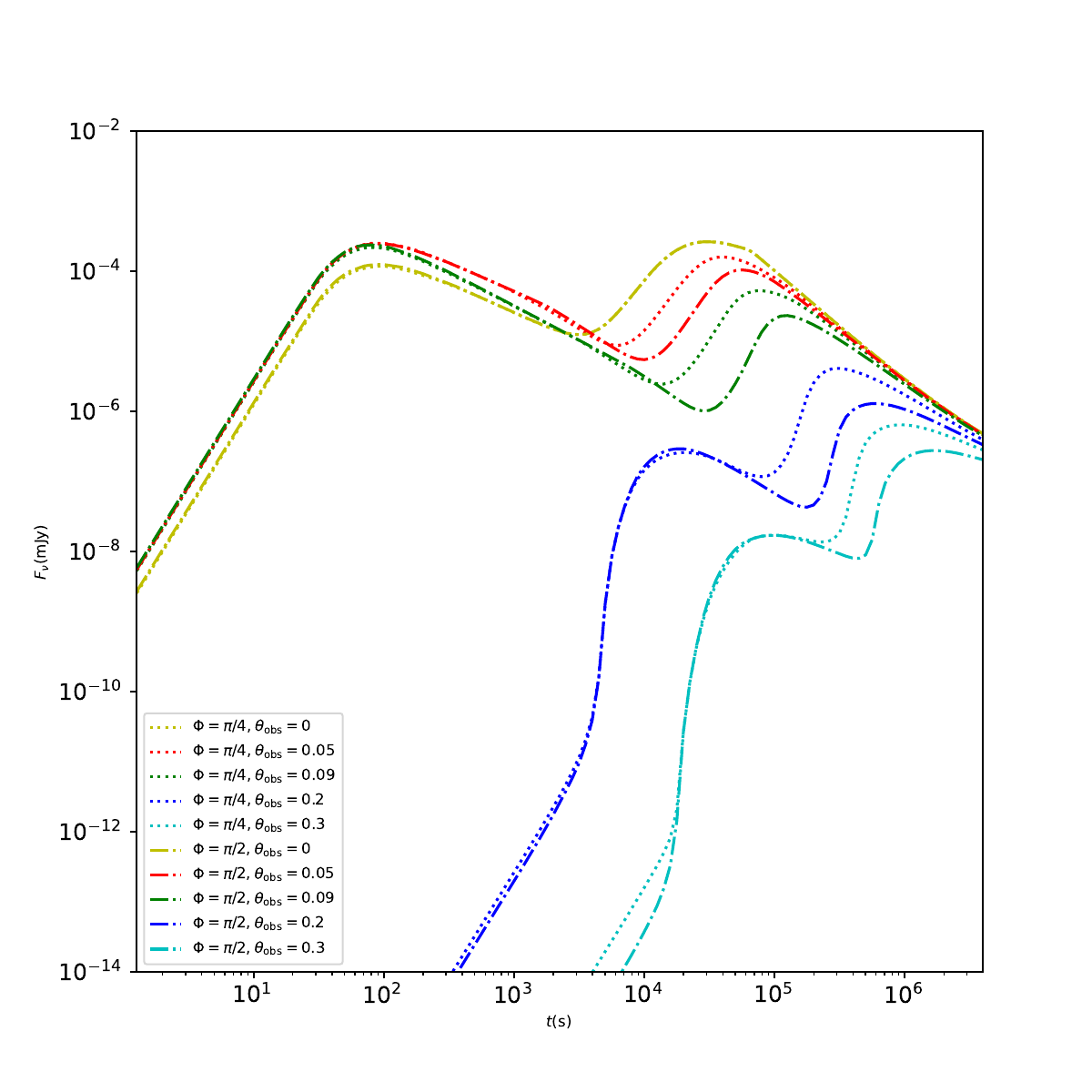}
		\label{Phito0.1s}}
	\subfigure[Semi-analytic approach]{\includegraphics[width=8.5cm]{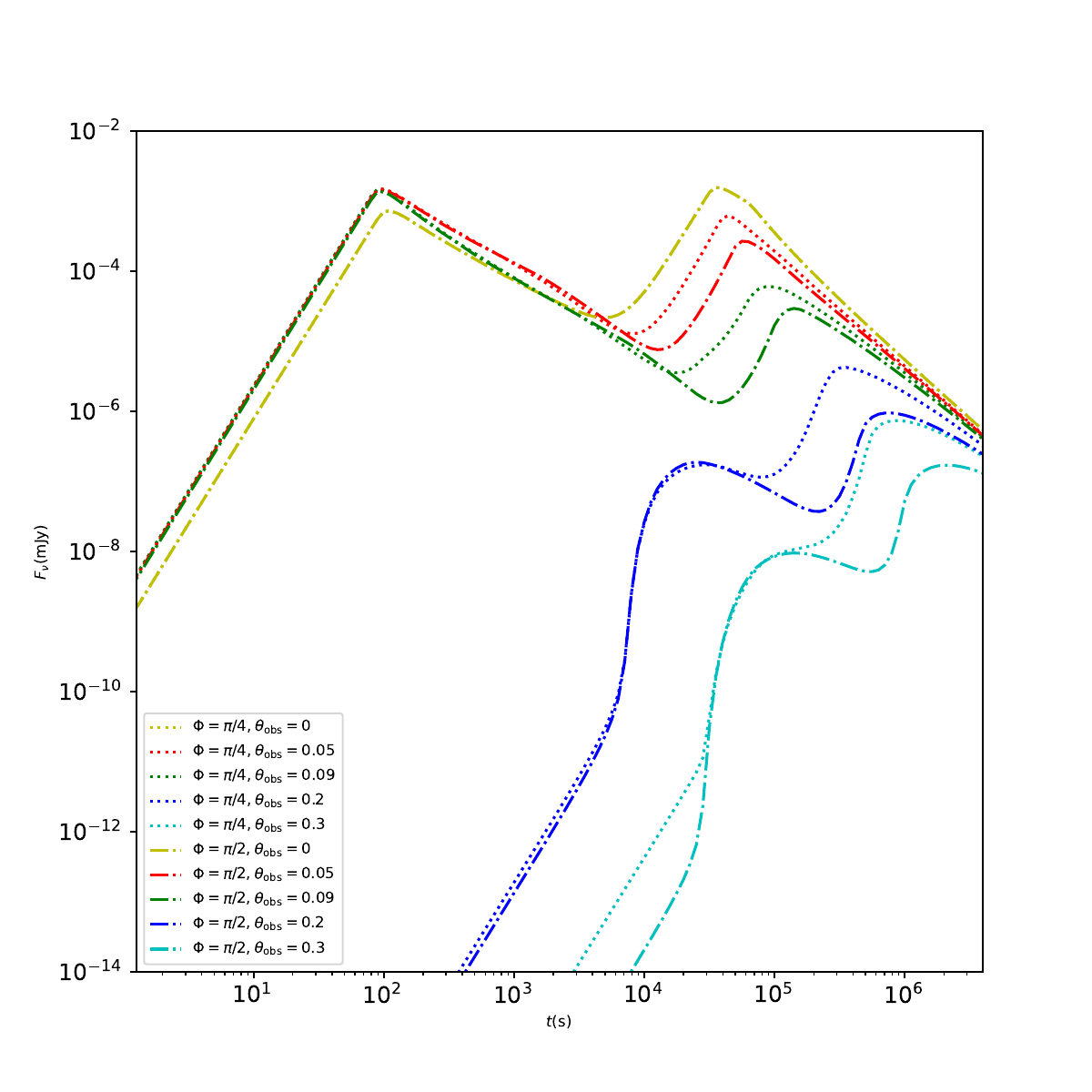}
		\label{Phito0.1n}}\\
        \subfigure[Numerical approach]
        {\includegraphics[width=8.5cm]{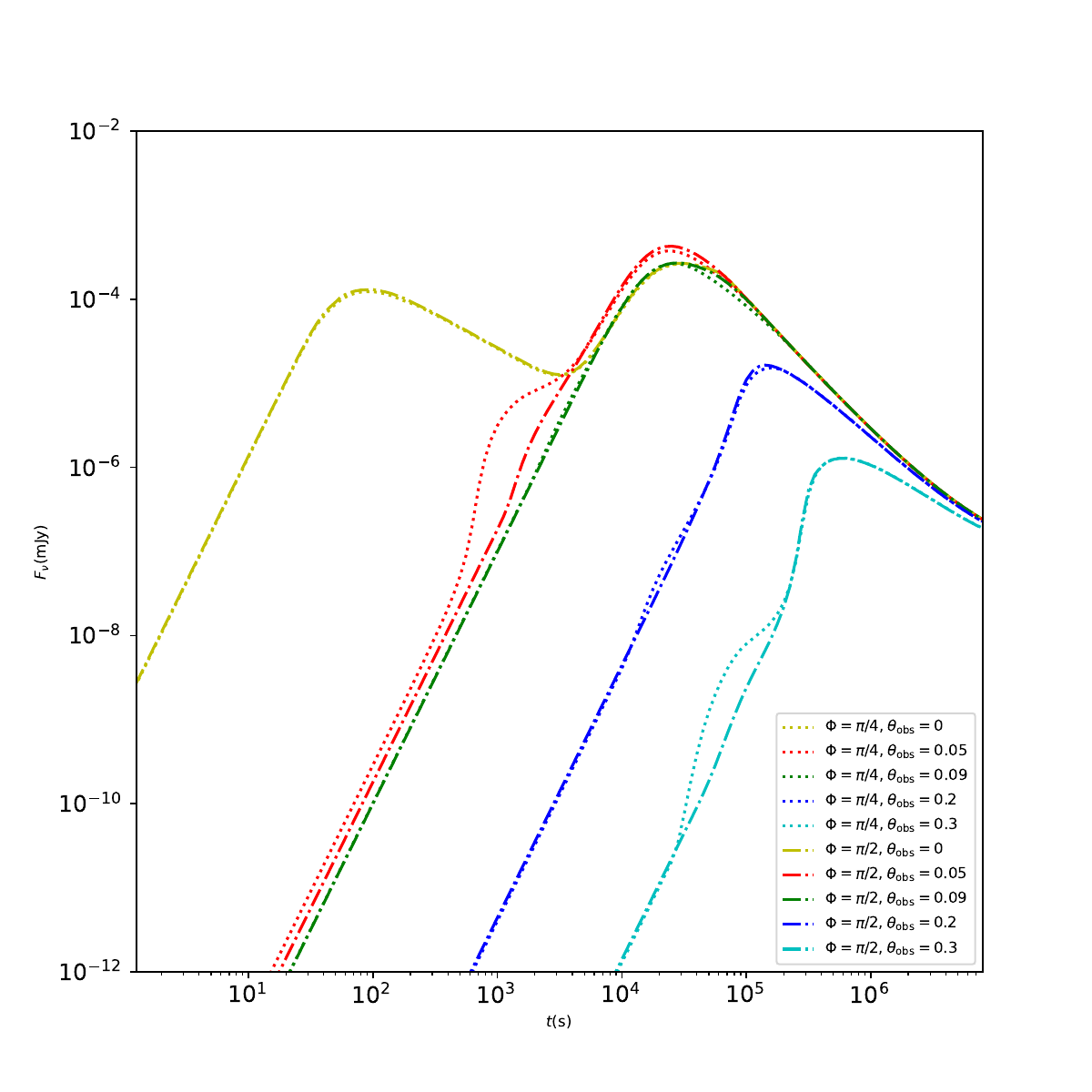}
		\label{Phito0.1sf}}
	\subfigure[Semi-analytic approach]{\includegraphics[width=8.5cm]{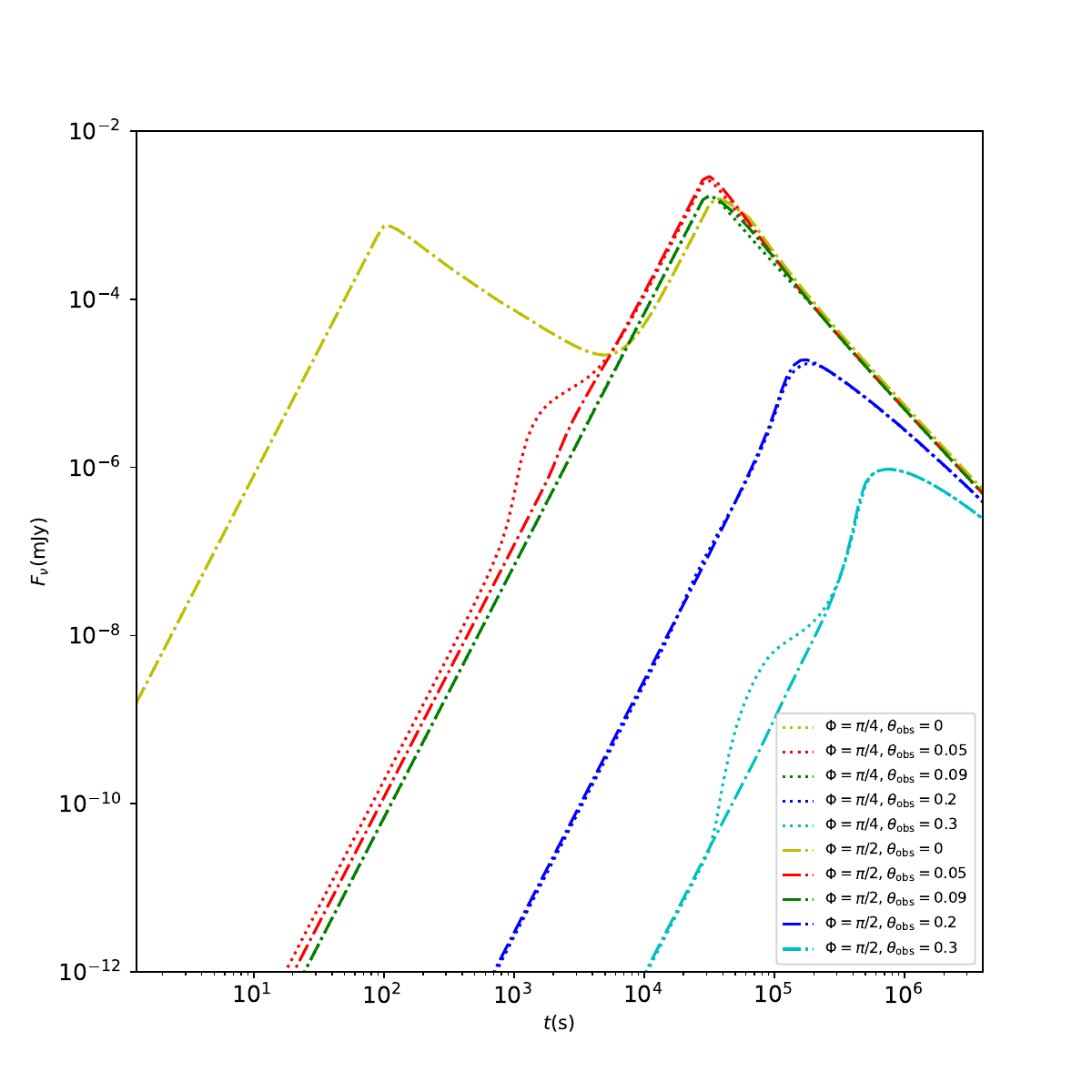}
		\label{Phito0.1nf}}\\
	\caption{The light curves of afterglows with varying $\Phi$ and $\theta_{\text{obs}}$. Different color represent different values of $\theta_{\text{obs}}$, and different styles of lines represent different values of $\Phi$.}
	\label{PHI}
\end{figure*}

In this scenario, the shape of the light curve depends not only on the physical parameters of two elements, but also on the values of $\Phi$ and $\theta_{\text{obs}}$.  Figure \ref{PHI} shows the light curves of afterglows with varying $\Phi$ and $\theta_{\text{obs}}$. We adopt a fixed value of $\theta_{\text{j}}=0.1$, whilst allowing $\theta_{\text{obs}}$ to vary between $0$ and $0.3$. And we compare the cases $\Phi=\pi/4$ and $\Phi=\pi/2$. We set the initial Lorentz factors as $\gamma_{01}=100$ and $\gamma_{02}=20$, and the isotropic energy as $E_{\text{iso},1}=10^{50} \text{ergs}$ and $E_{\text{iso},2}=10^{52} \text{ergs}$. Other parameters are identical to those in Figure \ref{gamma}. 

In figure \ref{Phito0.1s} and \ref{Phito0.1n}, we assume the LOS is more inclined towards the element with a larger Lorentz factor. In this case, when $\theta_{\text{obs}}<\theta_{\text{j}}$, the light curve contains two distinct peaks, with the first peak being less affected by changes in $\Phi$ and $\theta_{\text{obs}}$. The second peak will appear delayed and weakened as $\Phi$ and $\theta_{\text{obs}}$ increases. On the other hand, when $\theta_{\text{obs}}>\theta_{\text{j}}$, both peaks will appear delayed and weakened as $\theta_{\text{obs}}$ increases. Moreover, the variation of $\Phi$ can further affect the second peak.

Figures \ref{Phito0.1sf} and \ref{Phito0.1nf} show the light curves when the LOS leans towards to the element with lower Lorentz factor. In this case, when $\theta_{\text{obs}}<\theta_{\text{j}}$, the light curve often only contains one peak that appears relatively later, and the original first peak will become a small bump in the rising stage, which will gradually disappear with the further increase of $\Phi$. On the other hand, when $\theta_{\text{obs}}>\theta_{\text{j}}$, the light curve will be dominated by the contribution from the element with the lower Lorentz factor. The other element may or may not produce a bump in the rising stage, depending on the specific energy and velocity ratios between the two elements.

\subsection{More than 2 elements in the jet}

For more complex asymmetric structures, the jet may be divided into multiple elements with $N>2$. These individual elements exhibit differences in both $\gamma_0$ and $E_{\rm iso}$, which consequently leads to time-varying afterglow radiation observable by the observer. In principle, each element could produce a distinct peak, with the timing and magnitude of the peak dependent on the energy, velocity, and LOS of the corresponding element. The superposition of multiple radiation components can result in various intriguing types of light curves:
\begin{itemize}
    \item when the LOS is aligned with the axis of the jet and there are significant differences in the physical parameters of each element, the light curve may exhibit multiple distinct peaks;
    \item when the LOS is aligned with the axis of the jet and there are significant differences in the velocities of each element while the energy differences remain small, the light curve may exhibit a plateau;
    \item when the LOS is inclined towards the faster-moving elements, the peak of the light curve appears earlier, and in the later stage, there is a possibility of either a re-brightening or the absence of a re-brightening, depending on the energy magnitude of the slower-moving elements;
    \item when the LOS is inclined towards the slower-moving elements, the peak of the light curve appears later, and in the early rising phase, there is a possibility of encountering some fluctuations or not encountering any fluctuations, depending on the energy magnitude of the fast-moving elements;
    \item when the LOS is significantly larger than the jet opening angle, the peak of the light curve appears later. In the rising and falling phases, there is a possibility of encountering some fluctuations or not encountering any fluctuations, primarily determined by the elements closer to the LOS.
\end{itemize}

\begin{figure}
    \centering
    \includegraphics[width=8.5cm]{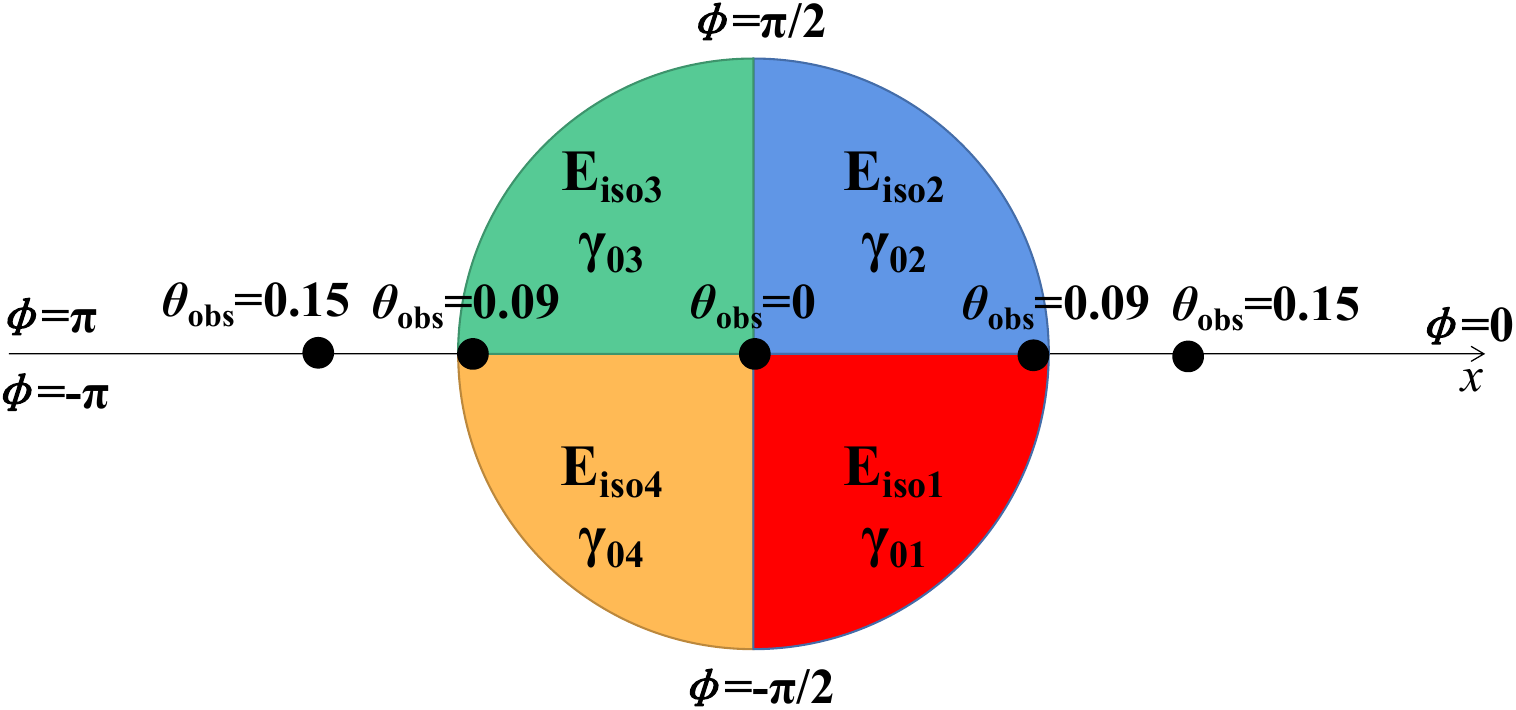}
    \caption{The schematic diagram illustrates the cross-sectional view of a four-element jet, where the interfaces of the jet are located either on the $x$-axis or are perpendicular to it. The various lines of sight with different angles of observation $\theta_{\text{obs}}$ are indicated on the $x$-axis for reference.}
    \label{3.3}
\end{figure}

\begin{figure*}
\centering
	\subfigure[Case I (numcrical approach)]{\includegraphics[width=8.5cm]{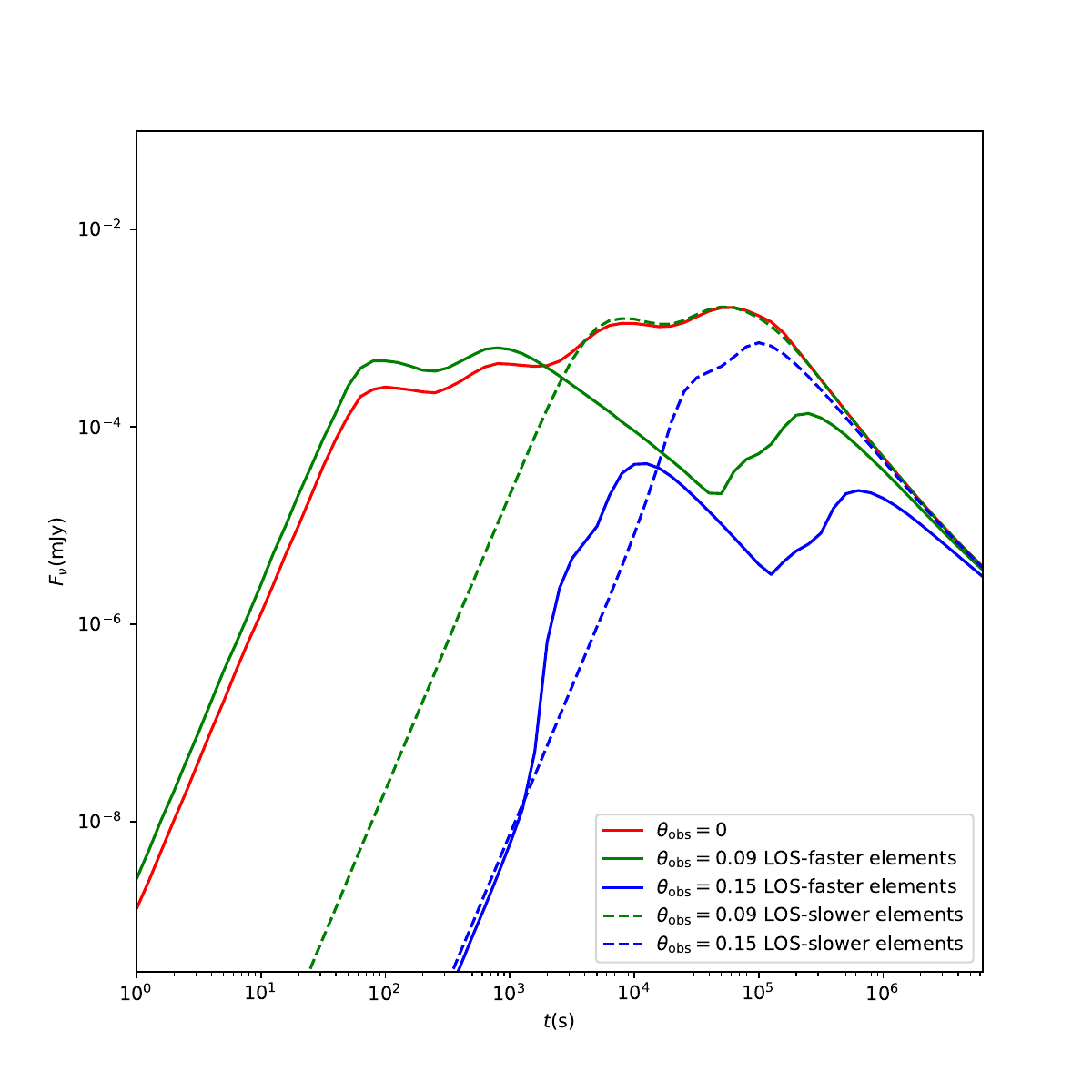}
		\label{4partsz}}
	\subfigure[Case I (semi-analytic approach)]{\includegraphics[width=8.5cm]{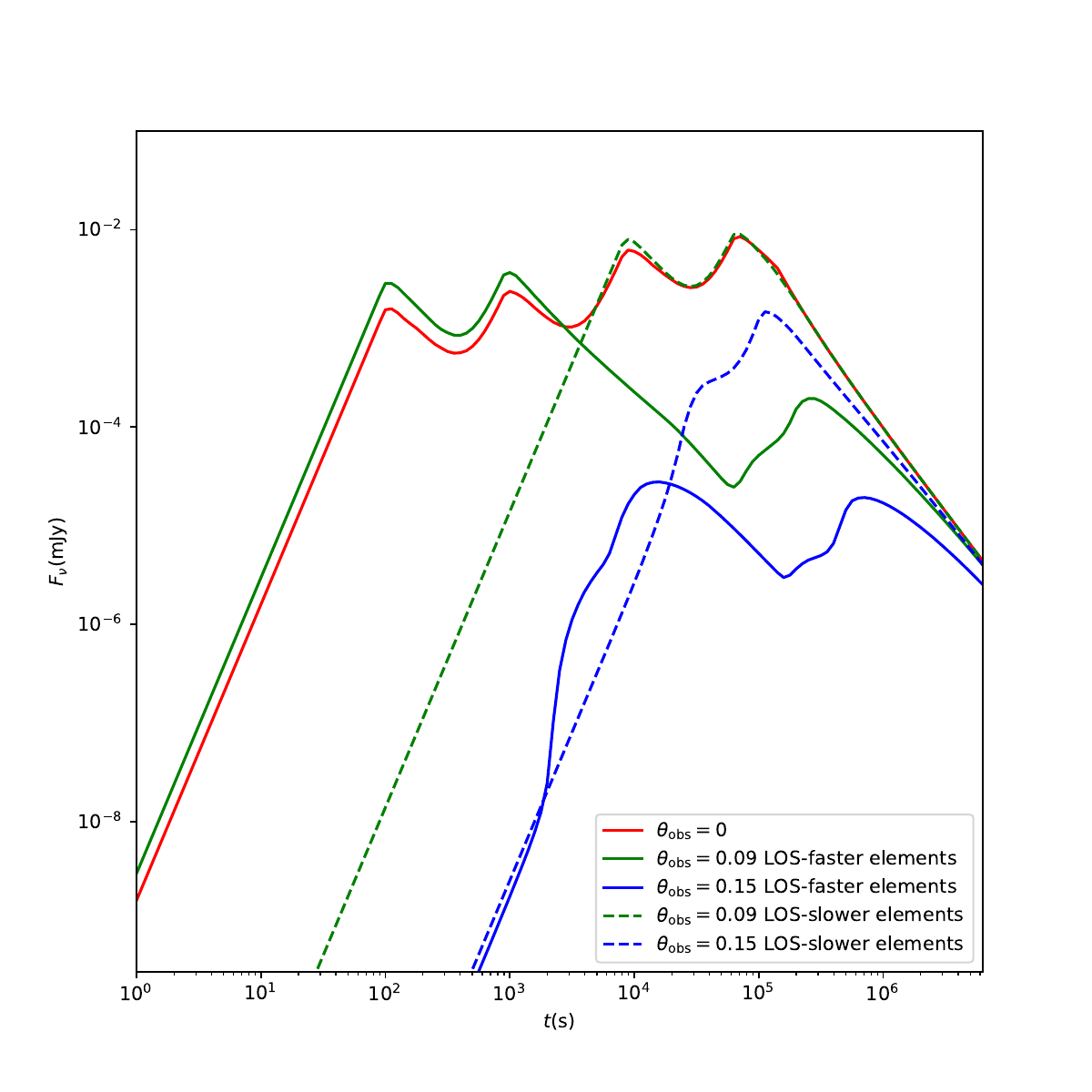}
		\label{4partjx}}
        \subfigure[Case II (numcrical approach)]{\includegraphics[width=8.5cm]{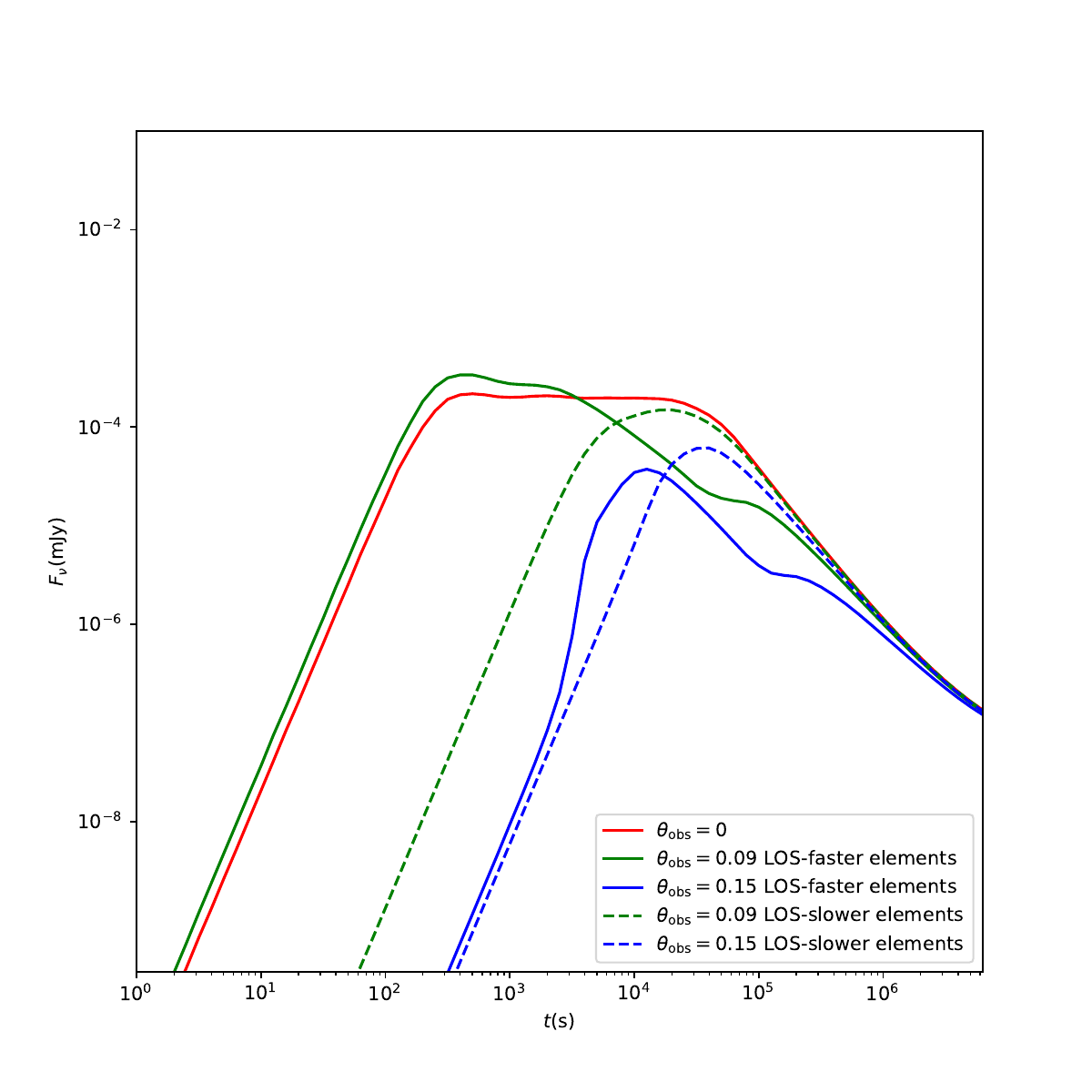}
		\label{4partsz2}}
	\subfigure[Case II (semi-analytic approach)]{\includegraphics[width=8.5cm]{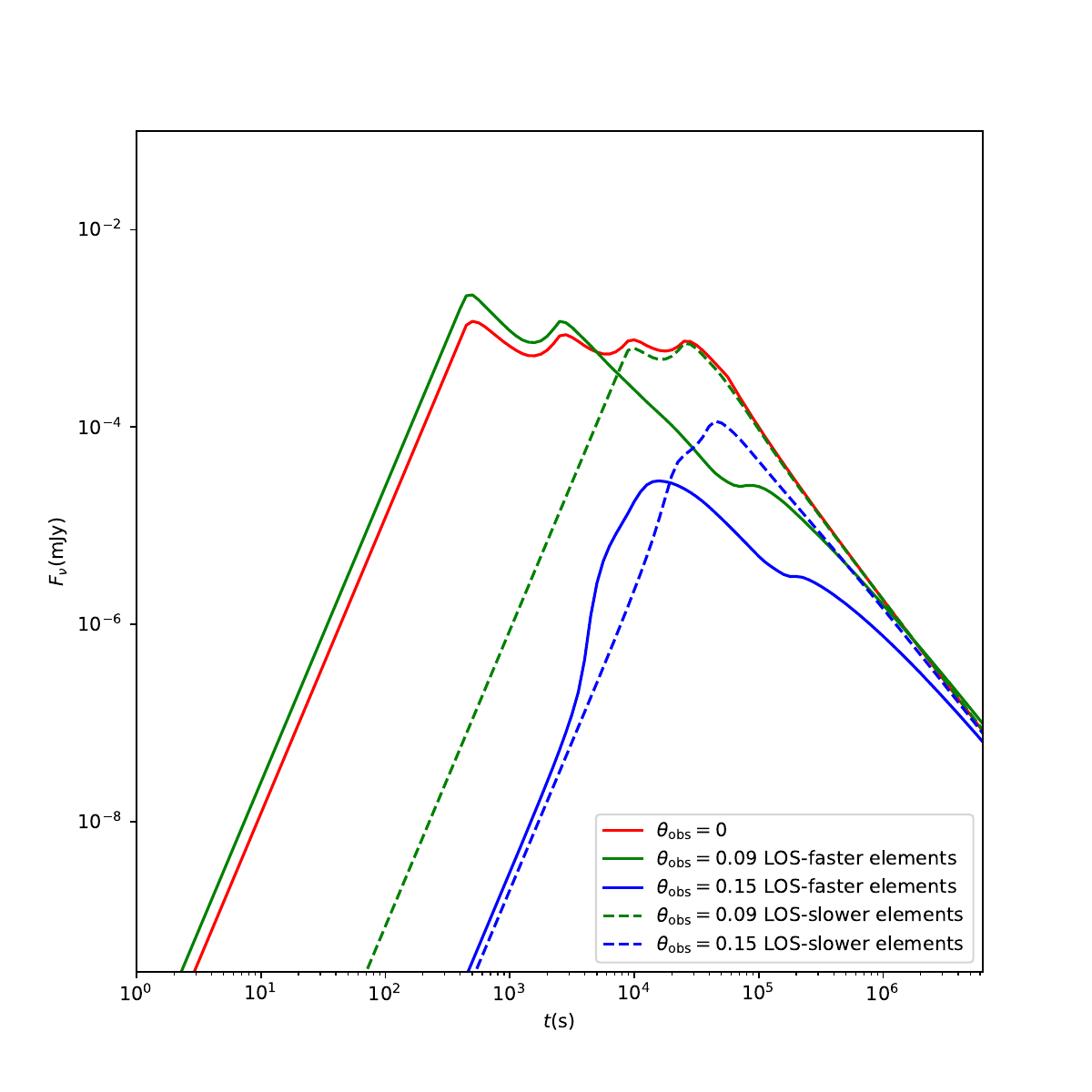}
		\label{4partjx2}}
	\caption{The light curves of afterglows for a four-element jet, with the parameters between each element exhibiting discernible differences (Case I) or inconspicuous differences (Case II). Different colors represent different $\theta_{\text{obs}}$ values. For $\theta_{\text{obs}}=0.09$ and $\theta_{\text{obs}}=0.15$, the solid and dashed lines indicate that the LOS is more inclined towards elements with higher Lorentz factors or lower Lorentz factors.}
	\label{4element}
\end{figure*}

As an example, we study a complicated structure with four elements. The schematic picture is shown in Figure \ref{3.3}. The four elements are equally distributed at different azimuth ranges, with different $\gamma_0$ and $E_{\text{iso}}$ values.  
Its structure is mathematically defined as, 
\begin{equation}
    \gamma_0=\begin{cases}
    \gamma_{01}&-\pi/2<\phi<0,\\
    \gamma_{02}&0<\phi<\pi/2,\\
    \gamma_{03}&\pi/2<\phi<\pi,\\
    \gamma_{04}&-\pi<\phi<-\pi/2,
    \end{cases}
    \label{g4}
\end{equation}
and
\begin{equation}
    E_{\text{iso}}=\begin{cases}
    E_{\text{iso}1}&-\pi/2<\phi<0,\\
    E_{\text{iso}2}&0<\phi<\pi/2,\\
    E_{\text{iso}3}&\pi/2<\phi<\pi,\\
    E_{\text{iso}4}&-\pi<\phi<-\pi/2.
    \end{cases}
    \label{E4}
\end{equation}
Figures \ref{4element} illustrate the afterglows from a four-element jet. For Case I (see figures \ref{4partsz} and \ref{4partjx}), we set the initial Lorentz factor of the four elements as $\gamma_{01}=20$, $\gamma_{02}=50$, $\gamma_{03}=75$, and $\gamma_{04}=110$ respectively. And their isotropic energies are $E_{\text{iso},1}=10^{53} \text{ergs}$, $E_{\text{iso},2}=10^{52} \text{ergs}$, $E_{\text{iso},3}=10^{51.3} \text{ergs}$ and $E_{\text{iso},4}=10^{50} \text{ergs}$ respectively. And $\theta_{\text{j}}=0.1$. Other parameters are consistent with those in section \ref{subsec:two-element-alinged}. We have analyzed the situations from five different LOS: 1) when $\theta_{\text{obs}}=0$, the jet exhibits a characteristic light curve with four distinct peaks; 2) when $\theta_{\text{obs}}=0.09,~\phi=\pi$ (i.e. the LOS is within the jet and more inclined towards elements with higher Lorentz factors), the light curve exhibits two prominent main peaks, followed by a weaker re-brightening period with noticeable fluctuations in its slope; 3) when $\theta_{\text{obs}}=0.15,~\phi=\pi$ (i.e. the LOS is outside the jet and more inclined towards elements with higher Lorentz factors), the light curve reveals two clear peaks, both exhibiting changes in slope during the rising phase; 4) when $\theta_{\text{obs}}=0.09,~\phi=0$ (i.e. the LOS is inside the jet and more inclined towards elements with lower Lorentz factors), the light curve displays two peaks at relatively late time; 5) when $\theta_{\text{obs}}=0.15,~\phi=0$ (i.e. the LOS is outside the jet and more inclined towards elements with lower Lorentz factors), there is only one peak at late time with wiggling feature during the rising phase. 

For Case II (see figures \ref{4partsz2} and \ref{4partjx2}), we selected four elements with Lorentz factors $\gamma_{01}=20$, $\gamma_{02}=27$, $\gamma_{03}=40$, and $\gamma_{04}=70$, and isotropic energies $E_{\text{iso},1}=10^{51.8} \text{ergs}$, $E_{\text{iso},2}=10^{51.5} \text{ergs}$, $E_{\text{iso},3}=10^{51.2} \text{ergs}$, and $E_{\text{iso},4}=10^{50.9} \text{ergs}$. The differences in Lorentz factor and isotropic energy between these elements are not significant. Compared to Case I, when $\theta_{\text{obs}}=0$, the light curve does not exhibit clear four peaks, but rather a flattened shape near the peak. When $\theta_{\text{obs}}=0.09$ and $\theta_{\text{obs}}=0.15$, the impact of the observation angle on the light curve resembles that of Case I. Nonetheless, due to the marginal disparity in the physical parameters of each component, the perturbations in the light curve are comparatively attenuated. Consequently, certain peaks have been mitigated to minor fluctuations or have even ceased to manifest. 

\section{Conclusions and Discussions}
\label{DC}

In this study, we intend to conduct a first step analysis of the potential characteristics of gamma-ray burst afterglows within the framework of non-axisymmetric structured jets, where the physical parameters vary along the azimuthal direction. To accomplish this, we simplify the profile of the asymmetric jet as a step function of the azimuth $\phi$, dividing the entire jet into $N$ individual elements. Each element is considered to be uniform and independent. The total light curve of the afterglow, driven by the entire jet, is approximately estimated by superimposing the light curve associated with each individual element.

By considering specific cases with $N=2$ and $N=4$, we find that the velocity, energy, and line-of-sight direction of each element can significantly impact the behavior of the overall light curve. The radiative contributions from multiple elements may result in the appearance of multiple distinct peaks or plateaus in the light curve. Or in some cases only a small number of peaks, but there are clear signs of fluctuations in the rising and declining segments of each peak. 

It is worth noting that if some simple variations appear in the GRB afterglow light curve, such as a single re-brightening feature, they could also potentially be generated by axisymmetric structured jet \citep[e.g., two component jet model;][]{Huang2004ApJ,Peng2005ApJ,Wu2005MNRAS,Beniamini2020MNRAS}. However, if the light curve shows more intricate patterns, such as multiple peaks or plateaus, the explanation relying on axisymmetric structured jets becomes challenging, since it is generally difficult for axisymmetric structures to exhibit a variety of discrete energy/velocity distribution patterns. The accumulated dataset of GRB optical afterglow currently exhibits a significant number of sources displaying indications of multiple peak and plateau configurations \citep{Li2012}. In the future, detailed fitting of our model to these sources holds the potential to enhance our comprehension of the structural characteristics of GRB jets. On the other hand, in the future, conducting additional numerical simulations, similar to \cite{Lamb2022Univ}, that incorporate the effects of jet precession or non-uniform jet dissipation, will contribute to the verification of the physical origin of non-axisymmetric jets.

\section*{Acknowledgements}

This work is supported by the National Natural Science Foundation of China (Projects 12021003), and the National SKA Program of China (2022SKA0130100). SA acknowledges the China Postdoctoral Science Foundation (2023M732713).

%%%%%%%%%%%%%%%%%%%%%%%%%%%%%%%%%%%%%%%%%%%%%%%%%%
\section*{Data Availability}

No new data were generated or analysed in support of this research.



\bibliographystyle{mnras}
\bibliography{example} % if your bibtex file is called example.bib

%\begin{thebibliography}{99}
%\end{thebibliography}

%%%%%%%%%%%%%%%%%%%%%%%%%%%%%%%%%%%%%%%%%%%%%%%%%%

%%%%%%%%%%%%%%%%% APPENDICES %%%%%%%%%%%%%%%%%%%%%

%\appendix


%If you want to present additional material which would interrupt the flow of the main paper, it can be placed in an Appendix which appears after the list of references.

%%%%%%%%%%%%%%%%%%%%%%%%%%%%%%%%%%%%%%%%%%%%%%%%%%

% Don't change these lines
\bsp	% typesetting comment
\label{lastpage}
\end{document}